\documentclass[twocolumn,showpacs,preprintnumbers,amsmath,amssymb]{revtex4-1}

\usepackage{graphicx}
\usepackage{dcolumn}
\usepackage{bm}

\newcommand{\qo}[1]{``#1''}                  
\newcommand{\ket}[1]{|#1\rangle}             

\begin{document}

\preprint{APS/123-QED}

\title{Polarization-controlled evolution of light transverse modes and associated Pancharatnam geometric phase in orbital angular momentum}

\author{Ebrahim Karimi$^{1,2}$}
\author{Sergei Slussarenko$^{1,3}$}%
\author{Bruno Piccirillo$^{1,3}$}%
\author{Lorenzo Marrucci$^{1,2}$}%
\author{Enrico Santamato$^{1,3}$}%
\affiliation{%
$^{1}$ Dipartimento di Scienze Fisiche, Universit\`{a} di Napoli \qo{Federico II}, Complesso di Monte S. Angelo, 80126 Napoli, Italy \\
$^{2}$ CNR-INFM Coherentia, Complesso di Monte S. Angelo, 80126 Napoli, Italy \\
$^{3}$ Consorzio Nazionale Interuniversitario per le Scienze Fisiche della Materia, Napoli, Italy
}%

\date{\today}

\begin{abstract}
We present an easy, efficient and fast method to generate arbitrary linear combinations of light orbital angular momentum eigenstates $\ell=\pm 2$ starting from a linearly polarized TEM$_{00}$ laser beam. The method exploits the spin-to-orbital angular momentum conversion capability of a liquid-crystal-based $q$-plate and a Dove prism inserted in a Sagnac polarizing interferometer. The nominal generation efficiency is 100\%, being limited only by reflection and scattering losses in the optical components. When closed paths are followed on the polarization Poincar\'{e} sphere of the input beam, the associated Pancharatnam geometric phase is transferred unchanged to the orbital angular momentum state of the output beam.
\end{abstract}

\maketitle

\section{Introduction}
In recent years a great deal of effort has been devoted to the creation and manipulation of paraxial optical beams which are eigenstates of the orbital angular momentum (OAM)~\cite{frankearnold08}. This kind of beams are promising for a wide range of applications, both in the classical and quantum regimes of light, thanks to the multi-dimensionality of their space~\cite{gibson04,mair01,molinaterriza07,nagali09a}. In many practical cases, the OAM space is however restricted to the $o_\ell$ subspace spanned by a pair of opposite OAM eigenvalues $\pm\ell$. This bidimensional optical subspace is then isomorphic to the standard polarization space, that is the space of the spin angular momentum (SAM) of light. A standard geometric representation of any polarization (or SAM) state of light is provided by the well known Poincar\'{e} sphere. In particular, the spin $s=\pm 1$ eigenvalues are usually mapped onto the poles of Poincar\'{e} sphere and correspond to left- and right-handed circular polarizations, while their equal-weight linear combinations are mapped along the equator circle and correspond to differently oriented linear polarizations. The other points on the sphere describe arbitrary elliptical polarizations. Analogously, any state in a given $o_\ell$ subspace can be represented as a point on a OAM Poincar\'{e} sphere~\cite{padgett99}. The OAM eigenvalues $\pm\ell$ may again be mapped onto the poles of this sphere and correspond to Laguerre-Gauss (LG$_\ell$) transverse modes, while equal-weight linear combinations are mapped along the equator circle and correspond to differently oriented Hermite-Gauss (HG$_\ell$) modes~\cite{footnote1}. By this geometrical representation, an one-to-one correspondence is established between the SAM space and $o_\ell$ subspace, for any $\ell$. In the following, we label the axis joining the right ($R$) and left ($L$) circular polarizations states on the Poincar\'{e} sphere as the $z$-axis, and the axis joining the vertical ($V$) and horizontal ($H$) polarization states as the $x$-axis. The photon SAM can be manipulated easily by polarizers and birefringent plates. It is well known that sequences of quarter-wave plates (QW) and half-wave plates (HW) oriented at suitable angles can change any given polarization state into another state at will~\cite{simon90,bhandari90}.
Moreover, electro-optical devices can be used to make such light polarization manipulation very fast. No so simple and fast devices are available for manipulating the light OAM. Cylindrical lens converters and Dove prisms can simulate the behavior of QW and HW wave plates in the $o_\ell$ space~\cite{padgett02}, but these devices are difficult to be aligned and cannot be used to manipulate OAM very quickly. Moving along continuous paths on the OAM Poincar\'{e} sphere by means of these devices would require very careful control and precise mechanics. Recently, a new device, the $q$-plate (QP) has been introduced which is able to transfer the SAM state of the beam to the $o_2$ subspace~\cite{marrucci06,marrucci06a}. The QP is a birefringent plate made of liquid crystals (LC) azimuthally aligned so that the local optical axis has a topological charge $q=1$ in the center of the plate. When a left-(right-)circularly polarized light beam traverses the QP, a topological charge $2q$ is transferred into its phase and the beam thus acquires an OAM $\ell=2$ ($\ell=-2$). Being cylindrically symmetric, the QP cannot exchange angular momentum and the OAM gained by the beam is balanced by a corresponding variation of its SAM. The net effect is a spin-to-orbital angular momentum conversion (STOC). The STOC process is complete only if the QP is ``tuned'', i.e. only if its birefringent retardation is half wavelength. Tuning of LC-based QPs can be achieved changing temperature~\cite{karimi09a},  pressure on the $q$-plate, or applying an external electrical field. The STOC process is coherent and it can be used to transfer complete qubit information from the SAM to the OAM degree of freedom at the single photon level~\cite{nagali09}. The STOC efficiency, i.e. the fraction of photons (or optical energy) that is actually converted can ideally reach 100\%, when reflection and scattering losses are neglected~\cite{karimi09b}. In a previous work, we showed that it is possible to achieve efficiencies exceeding 90\% by controlling the QP temperature~\cite{karimi09a}. A very appealing use of the STOC process is that of exploiting the easy and fast control that we have on the light polarization degree of freedom for controlling the OAM degree of freedom. By using STOC, cylindrical lens mode converters and Dove prisms can be conveniently replaced with birefringent plates and electro-optical cells.

The aim of this work is to demonstrate the easy and efficient control on the OAM of a light beam that is attainable via the STOC process. Arbitrary and continuously controllable linear combinations of LG$_2$ modes have been generated in a very simple way and with efficiency exceeding 90\% by manipulating the input beam polarization. Our method provides a highly efficient tool that, for example, can be advantageously used to create photonic qudits involving both SAM and OAM degrees of freedom, thus enlarging the amount of information carried by single photons. The use of qudits instead of  qubits may for example lead to simplifying quantum computations~\cite{muthukrishnan00,lanyon08} and improving quantum cryptography~\cite{bergman08}. Having a versatile and fast way to handle single photon OAM, besides, will help implementing quantum logic schemes based on the single-photon multi-qubit (SPMQ) coding, where several qubits can be encoded in different transverse modes of one photon~\cite{vaidman98}.
\begin{figure}[!htbp]
\centerline{\includegraphics[draft=false,width=9cm]{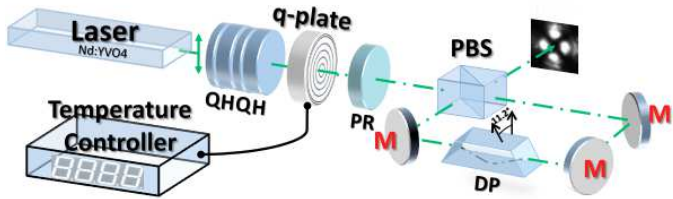}}
\caption{\label{fig:1} (Color online) Experimental setup for generating polarization-controlled linear combinations of LG$_2$ beams. Legend: QHQH - set of waveplates to control the beam polarization; PR - polarization rotator; PBS - polarizing beam-splitter; DP - Dove prism; M - mirror.}
\end{figure}
\section{Experimental setup}
Our experimental setup is shown in Fig.~\ref{fig:1}. The polarization of the light beam entering the QP is controlled by the QHQH waveplate sequence QW(90\ensuremath{^\circ})HW($-\gamma/4$)QW(0\ensuremath{^\circ})HW(90\ensuremath{^\circ}+$\delta/4$), where in parentheses there are the orientation angles of each plate counted from the horizontal plane. As it can be easily shown, this sequence of waveplates applies to the input polarization state first an SU(2) transformation consisting of a rotation of angle $\gamma$ around the $y$-axis of the SAM Poincar\'{e} sphere and then a rotation of an angle $\delta$ around the $z$-axis. After this QHQH set, we inserted a tuned QP, a polarization rotator (PR) of 45\ensuremath{^\circ}~\cite{footnote2} and a Sagnac Polarizing Interferometer (SPI). In the SPI, a Dove prism rotated by 11.25\ensuremath{^\circ} from the horizontal plane was inserted. The QP, PR, and SPI are the heart of our apparatus, because they realize the required mapping~\cite{footnote3}
\begin{equation}\label{eq:mapping}
  (\alpha\ket{L}+\beta\ket{R})\ket{0}\rightarrow(\alpha\ket{2}+\beta\ket{-2})\ket{D}
\end{equation}
where $\ket{L},\ket{R},\ket{H},\ket{V},\ket{D},\ket{A}$ denote the right-circular, left-circular, horizontal, vertical, diagonal, and anti-diagonal polarization states, respectively, and $\ket{\ell}$, with integer $\ell$, denotes the OAM eigenstate with eigenvalue $\ell$. It is worth noting that our apparatus works in any basis. For example, it also realizes the mapping $(\alpha\ket{H}+\beta\ket{V})\ket{0}\rightarrow(\alpha\ket{h}+\beta\ket{v})\ket{D}$, where $\ket{h}$ and $\ket{v}$ denote the HG$_2$ modes corresponding to the linear polarization states $H$ and $V$ respectively. The detailed operation of our apparatus is the following. Up to a global phase factor, the action of a tuned QP on the elliptically polarized TEM$_{00}$ ($\ell=0$) input beam is given by
\begin{eqnarray}\label{eq:qp}
    \lefteqn{(\alpha\ket{L}+\beta\ket{R})\ket{0}\overset{\widehat{\mathrm{QP}}}{\rightarrow}
    \alpha\ket{R,2}+\beta\ket{L,-2}}  \nonumber \\
    \hspace{-1cm}=&&\hspace{-0.5cm}\frac{1}{\sqrt{2}}[\ket{H}(\alpha\ket{2}+\beta\ket{-2})-
    i\ket{V}(\alpha\ket{2}-\beta\ket{-2})]
\end{eqnarray}
The radial modes are factorized out and can be omitted~\cite{footnote4}. From Eq.~(\ref{eq:qp}) we see that insertion of a linear polarizer after the QP would already select the desired linear combination of LG$_{2}$ and LG$_{-2}$ modes (or $\ket{2}$ and $\ket{-2}$ states), but this would also reduce the maximum conversion efficiency to 50\%~\cite{nagali09}. The polarizing Sagnac interferometer scheme shown in Fig.~\ref{fig:1} allows one to increase the theoretical efficiency to 100\%. The Sagnac interferometer is made up of a polarizing beam splitter (PBS) and three mirrors (M). The $H$- and $V$-polarized components of the beam emerging from the $q$-plate are initially separated by the PBS and travel through the interferometer in opposite directions until they are recombined on exit by the same PBS. Equal optical paths of the counter-propagating beams render this interferometer particularly noise-insensitive, thus removing the need for active control of the interferometer length~\cite{fiorentino04}. The reflection in the Dove prism tilted at angle $\theta$ adds a phase factor $e^{2i\ell\theta}$ to the OAM eigenstate $\ket{\ell}$ and changes $\ket{\ell}$ into $\ket{-\ell}$. In our case, moreover, because of the counter-propagation, the $H$-polarized beam sees the Dove prism tilted at angle $\theta$ and the $V$-polarized beam sees the Dove prism tilted at angle $-\theta$. In order to calculate the overall effect of the Sagnac interferometer we must add to the action of the Dove prism the action on the OAM value of the three mirrors and PBS. The mirrors and PBS add an odd number of reflections to both the counterpropagating beams, so that the final value $\ell$ of the OAM after the Sagnac interferometer is the same as in the imput beam. With the substitutions $\ket{H,\ell}\rightarrow e^{2i\ell\theta}\ket{H,\ell}$, $\ket{V,\ell}\rightarrow e^{-2i\ell\theta}\ket{V,\ell}$, and setting $\theta=\pi/16$ in Eq.~(\ref{eq:qp}) we obtain, up to a phase factor, the state after the Sagnac interferometer as
\begin{equation}\label{eq:psiout}
   (2) \rightarrow \psi_{out}=\ket{D}(\alpha\ket{2}-i\beta\ket{-2}),
\end{equation}
We see that the resulting beam is fully polarized, and that its OAM content is a linear combination of LG$_{\pm 2}$ modes with coefficients uniquely related to $\alpha$ and $\beta$. Thus, we succeeded to realize one-to-one mapping of the input SAM state onto the output OAM state, but not yet the wanted one, given by Eq.~(\ref{eq:mapping}). The phase difference of $\pi/2$ between the two terms on the right of Eq.~(\ref{eq:psiout}) must be eliminated. This is accomplished by means of the PR located before the PSI, which introduces a retardation of $\pi/2$ between the circular polarization
components of the input beam.
\section{SAM-to-OAM mapping and detection of Pancharatnam geometric phase}
To show the flexibility of our apparatus in manipulating the light OAM, we performed a set of measurements in which we slowly modulated the polarization of the input TEM$_{00}$ beam in order to drive the state of the output beam along a predetermined trajectory on the OAM Poincar\'{e} sphere. In this way, arbitrary states in the $o_2$ subspace were easily and continuously generated starting from a TEM$_{00}$ laser beam. The power conversion efficiency from TEM$_{00}$ $H$-polarization to $o_2$ modes was found to exceed 90\% for all $o_2$ modes. This efficiency is larger than the maximum typically obtainable ($\simeq 70$\%) with blazed holograms. In our experiments, we measured the OAM content of the output beam in several points on the Poincar\'{e} sphere by recording the intensity profile and the pattern obtained by interference with a $D$-polarized TEM$_{00}$ mode as reference beam. For the sake of brevity, the interferometric apparatus adopted to record such interference patterns has not been reported in Fig.~\ref{fig:1}. The results are shown in Figs.~{\ref{fig:2}-\ref{fig:5}.
\begin{figure}[!htbp]
\centerline{\includegraphics[draft=false,width=9cm]{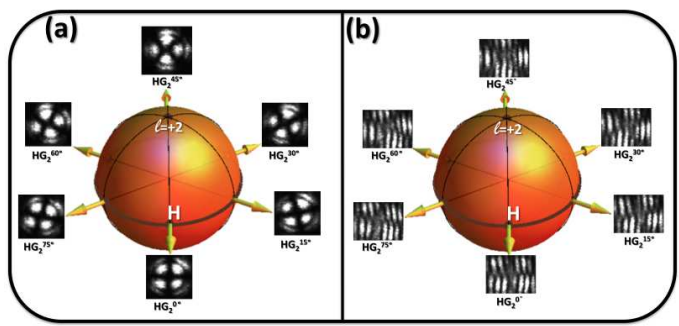}}
\caption{\label{fig:2} (Color online) Trajectory along the equator of the Poincar\'{e} sphere. H is the starting and ending point of the closed path. a) Intensity profiles of generated beams corresponding to differently rotated HG$_2$ modes. (b) Corresponding interference patterns with a $D$-polarized TEM$_{00}$ reference beam. As the state travels along the equator both the intensity and interferogram patterns rotate through an angle between $0\ensuremath{^\circ}$ and $90\ensuremath{^\circ}$.}
\end{figure}
\begin{figure}[!htpb]
\centerline{\includegraphics[draft=false,width=7cm]{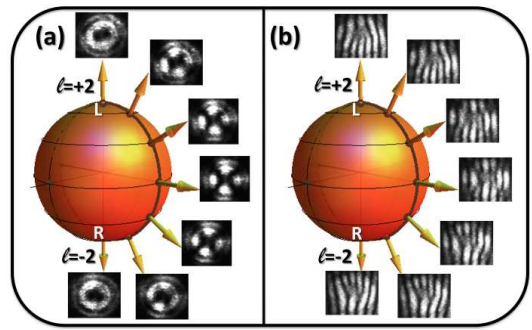}}
\caption{\label{fig:3} (Color online) Trajectory along a meridian of the Poincar\'{e} sphere.(a) Intensity profiles of generated beams corresponding to different linear combinations of LG$_{\pm2}$ modes. (b) Corresponding interference patterns with a $D$-polarized TEM$_{00}$ reference beam.}
\end{figure}

In the measurements we used a $532\,\hbox{nm}$ TEM$_{00}$ $H$-polarized laser beam and a home made LC $q$-plate thermally tuned to optimum STOC. The details about the $q$-plate and the tuning system have been reported elsewhere~\cite{karimi09a}. In Fig.~\ref{fig:2} there are shown the intensity profiles and interferograms of the HG$_2$ modes represented by the points located on the equator of the OAM Poincar\'{e} sphere. The modes shown in Fig.~\ref{fig:2} were generated fixing to $0\ensuremath{^\circ}$ the axis of the first HW-plate in the QHQH device and rotating the axis of the second HW-plate from $90\ensuremath{^\circ}$ to $180\ensuremath{^\circ}$. The input polarization state circulated along the equator of the SAM Poincar\'{e} sphere starting from $\ket{H}$, i.e. $\gamma=0$ and $0\leq\delta\le 2\pi$, denoted by the point $H$ on the sphere. The rest of the apparatus, implementing Eq.~(\ref{eq:mapping}), mapped the equator of the SAM sphere into the equator of the OAM sphere, which represents a continuous sequence of HG$_2$ modes whose transverse intensity and phase distributions linearly rotate clockwise from $0\ensuremath{^\circ}$ to $90\ensuremath{^\circ}$ (Fig.~\ref{fig:2}(a) and (b) respectively), the global phase being $\pi$-shifted over the closed path (see Fig.~\ref{fig:5}(a)). Such a phase shift over the cycle was inferred comparing the intensity profiles of the fringe patterns along the transverse direction of both the initial and the final states, represented in Fig.~\ref{fig:5}(a) by a dashed and a continuous lines respectively. Of course, when the initial state of the input beam is represented by a point located outside the equator, i.e. $\gamma\neq 0$ on the Poincar\'{e} sphere, the rotation of the axis of the second HW-plate in the device QHQH from $90\ensuremath{^\circ}$ to $180\ensuremath{^\circ}$ will yield into a variation of $\delta$ from 0 to $2\pi$. This amounts to drive the OAM state along a parallel on the Poincar\'{e} sphere. Analogously, the modes represented by the points located along a meridian trajectory on the OAM Poincar\'{e} sphere were produced fixing $\delta=0$ and driving $\gamma$ from $-\pi/2$ to $+\pi/2$ in the QHQH device, i.e. rotating the axis of the first HW-plate in the QHQH device from $+22.5\ensuremath{^\circ}$ to $-22.5\ensuremath{^\circ}$. In Fig.~\ref{fig:3}, it is shown the continuous passage from the LG$_2$ mode, i.e. doughnut intensity and down-fork interferogram, to HG$_{2}$ mode rotated by $45\ensuremath{^\circ}$, crossing the equator, till to LG$_{-2}$ mode, i.e. doughnut intensity again and up-fork interferogram. There were reported also the field distributions of the modes corresponding to points located along a meridian at intermediate positions between the poles and the equator.

On the grounds of the previous two experiments, it is clearly possible moving between two arbitrary OAM states passing through a continuous series of states represented by the points of suitable arcs of parallels and meridians. Furthermore, in principle, any path connecting two arbitrary points on the Poincar\'{e} sphere could be actually realized rotating simultaneously both the HW-plates in the QHQH device in order to approximate as better as possible the path by a polygonal chain consisting of small arcs of parallel and meridians. Finally we exploited this general feature in a simple case, to drive the OAM state along a closed path $L \rightarrow H \rightarrow A \rightarrow L$ on the Poincar\'{e} sphere (see Fig.~\ref{fig:4}), hinged on the north pole $L$. Therefore, starting from the state $\ket{2}$ (point $L$ in the figure), we rotated the first HW-plate from $+22.5\ensuremath{^\circ}$ to $0\ensuremath{^\circ}$, the second HW-plate being fixed at $90\ensuremath{^\circ}$. After this operation, the OAM state turns to be $\ket{h}$ (point $H$ in the figure). Then, the first HW-plate being fixed at $0\ensuremath{^\circ}$, the second HW-plate was rotated from $90\ensuremath{^\circ}$ to $112.5\ensuremath{^\circ}$, producing the OAM state $\ket{a}=(\ket{h}+\ket{v})/\sqrt{2}$ (point $A$ in the figure), which is the HG$_2$ mode corresponding to the linear polarization state $\ket{A}$. Finally the path was closed maintaining the second HW-plate fixed at $112.5\ensuremath{^\circ}$ and rotating the first one from $0\ensuremath{^\circ}$ to $+22.5\ensuremath{^\circ}$. In Fig.~\ref{fig:4} there are shown some intensity profiles and interferograms corresponding to the closed path considered. As inferred from the interferograms corresponding to the north pole in the initial and final states, closing the path results in the multiplication by a Pancharatnam geometric phase~\cite{pancharatnam}.
\begin{figure}[!htbp]
\centerline{\includegraphics[draft=false,width=9cm]{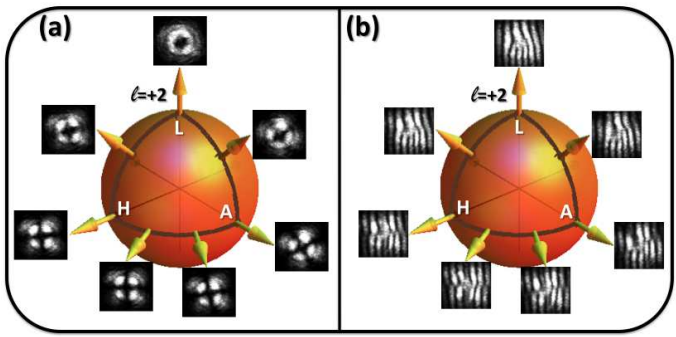}}
\caption{\label{fig:4} (Color online) A possible closed path over the OAM-Poincar\'{e} sphere. The path $L \rightarrow H \rightarrow A \rightarrow L$ starts and ends at the north pole, which associated to $\ket{+2}$. $H$ is the point associated to $\ket{h}$ and $A$ is the point associated to $\ket{a}$. (a) Intensity profiles of the generated beam at different points of the path. (b) Corresponding interference patterns with a TEM$_{00}$ $D$-polarized reference beam.}
\end{figure}
The observed fringe shift due to Pancharatnam geometric phase over a cycle is shown in Fig.~\ref{fig:5}(b). The global phase shift suffered by the state over the cycle $L \rightarrow H \rightarrow A \rightarrow L$ is $\pi/4$, as argued from Fig.~\ref{fig:5}(b). The intensity profiles of the fringe patterns along the transverse direction have been reported for both the initial and the final states simultaneously, caring about the maintenance of the same origin for the transverse coordinate in both the patterns for comparison. Dashed and continuous lines show the fringes related to the initial and final states, respectively.

In Fig.~\ref{fig:5} there have been reported both the phase shift related to the circulation along the equator (a) and that related to the circulation along the path $L \rightarrow H \rightarrow A \rightarrow L$ (b). As expected, in both cases it was found a phase shift equal to half the solid angle subtended by the path on the OAM Poincar\'{e} sphere~\cite{bhandari97}. The geometrical phase acquired by a light beam when the OAM state is moved along a closed path on the Poincar\'{e} sphere was observed some time ago~\cite{galvez03}. However, in this experiment the light OAM content was changed discontinuously having the beam pass through a sequence of fixed Dove prisms and cylindrical lens converters. In the present case, the beam OAM was changed adiabatically and the phase  continuously monitored along the path. Let us notice that the Pancharatnam geometric phase is already present when we close the state path on the SAM Poincar\'{e} sphere of the input beam. One of the issues of our experiment is therefore the demonstration that the STOC process is able to coherently transfer global phase shifts, such as geometric phase shifts, from SAM to OAM degree of freedom. Any path on SAM Poincar\'{e} sphere is mapped by Eq.~(\ref{eq:mapping}) into a corresponding path on linearly $D$-polarized OAM sphere. This feature allows for phase measurements with constant fringe visibility along any path, in particular along a closed path, which is impossible in experiments on Pancharatnam phase based on polarization only~\cite{bhandari97}.
\begin{figure}[!htbp]
\centerline{\includegraphics[draft=false,width=9cm]{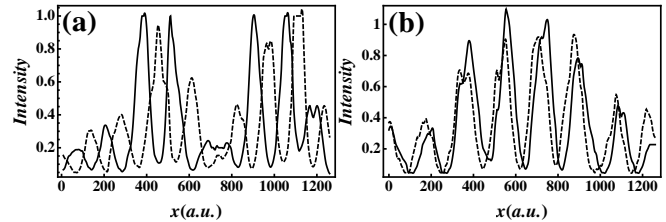}}
\caption{\label{fig:5} Interference patterns for two different closed trajectories on the Poincar\'{e} sphere. (a) Circular path
along the equator, as shown in Fig~\ref{fig:2}. In this case, there is a $\pi$ change in the phase when the path is closed. (b) Path shown in Fig~\ref{fig:4}. In this case there is a $\pi/4$ change in the phase when the path is closed. Dashed and continuous lines show the fringes of the initial and final states, respectively.}
\end{figure}
\section{Conclusions}
In conclusion, we exploited the STOC process in a temperature tuned $q$-plate to achieve an easy, fast and continuous control on the transverse modes of a laser beam. A particular Dove prism-based Sagnac polarizing interferometric configuration allowed us to generate with efficiency higher than 90\% arbitrary combinations of LG$_2$ modes in the $o_2$ subspace, by changing the polarization of the input TEM$_{00}$ linearly polarized laser beam. When closed paths are described on the SAM Poincar\'{e} sphere, identical closed path are described on the OAM Poincar\'{e} sphere and the resulting Pancharatnam geometric phase is transferred with no change from the SAM to the OAM degree of freedom. Our apparatus can work even in the single-photon regime, so we think it could be useful in many classical and quantum optics applications, where easy, fast and continuous manipulation of OAM is necessary.


\begin{thebibliography}{22}
\bibitem{footnote1}
By LG$_\ell$ modes we mean $\ell$-eigenstates of
light OAM resulting from linear combination of LG$_{p,\ell}$ modes with different
radial number $p$ and fixed azimuthal number $\ell$, i.e.  LG$_\ell=\sum_{p=0}^{\infty}c_{p}\hbox{LG}_{p,\ell}$.
Analogously, by HG$_\ell$ we mean equal-weight linear combinations of LG$_\ell$ and LG$_{-\ell}$,
i.e. HG$_\ell=\left(e^{i\ell\theta}\hbox{LG}_{\ell}+e^{-i\ell\theta}\hbox{LG}_{-\ell}\right)/\sqrt{2}$, where $\theta$ is the orientation of the HG$_{\ell}$ mode.
\bibitem{footnote2}
A polarization rotator may be implemented through an optically active medium or, as in the present experiment, through a couple of HW-plates with their axes rotated by 22.5\ensuremath{^\circ} with respect to each other.
\bibitem{footnote3}
Our apparatus produces constant $D$-linear polarization in the exit beam.
\end{thebibliography}

\end{document}